# System identification and closed-loop control of laser hot-wire directed energy deposition using the parameter-signature-property modeling scheme

M. Rahmani Dehaghani[1, a], Atieh Sahraeidolatkhaneh[b], Morgan Nilsen[b], Fredrik Sikström[b], Pouyan Sajadi[a], Yifan Tang[a], G. Gary Wang[a]

[a] School of Mechatronic Systems Engineering, Simon Fraser University, Burnaby, Canada
[b] Division of Production Systems, University West, Trollhattan, Sweden

**Abstract**

Hot-wire directed energy deposition using a laser beam (DED-LB/w) is a method of metal additive manufacturing (AM) that has benefits of high material utilization and deposition rate, but parts manufactured by DED-LB/w suffer from a substantial heat input and undesired surface finish. Hence, monitoring and controlling the process parameters and signatures during the deposition is crucial to ensure the quality of final part properties and geometries. This paper explores the dynamic modeling of the DED-LB/w process and introduces a parameter-signature-property modeling and control approach to enhance the quality of modeling and control of part properties that cannot be measured *in situ*. The study investigates different process parameters that influence the melt pool width (signature) and bead width (property) in single and multi-layer beads. The proposed modeling approach utilizes a parameter-signature model as $F_1$ and a signature-property model as $F_2$. Linear and nonlinear modeling approaches are compared to describe a dynamic relationship between process parameters and a process signature, the melt pool width ($F_1$). A fully connected artificial neural network is employed to model and predict the final part property, i.e., bead width, based on melt pool signatures ($F_2$). Finally, the effectiveness and usefulness of the proposed parameter-signature-property modeling is tested and verified by integrating the parameter-signature ($F_1$) and signature-property ($F_2$) models in the closed-loop control of the width of the part. Compared with the control loop with only $F_1$, the proposed method shows clear advantages and bears potential to be applied to control other part properties that cannot be directly measured or monitored *in situ*.

---
[1] Corresponding Author, Email: mostafa_rahmani_dehaghani@sfu.ca, ORCID ID: 0000-0003-4618-0302

# 1. Introduction

Laser wire directed energy deposition (DED-LB/w) technology involves a laser heat source to create a melt pool and adding filler materials in a wire form into the melt pool to produce metallic parts layer by layer [1]. DED-LB/w demonstrates an exceptional material utilization rate, approaching nearly 100% [2]. Moreover, this method allows for a rapid deposition rate, ensuring efficient and speedy fabrication [3]. Parts fabricated by DED-LB/w, however, require post processing machining and finishing to reach the desired surface finish and geometry. Furthermore, the substantial heat input during the process often leads to geometric deviations in the component caused by induced residual stresses. These deviations may result in manufacturing rejects or necessitate an oversized part [4]. Consequently, there is a growing interest in actively controlling and monitoring the dimensions and temperature of the part during the manufacturing process [5].

The melt pool size and temperature directly influence the geometry and shape of the deposited material. By precisely controlling these parameters, manufacturers can achieve the desired dimensions, layer thickness, and surface finish of the fabricated part. Most monitoring and control techniques focus on the characteristics of the melt pool, since the final quality of the part is highly correlated with the dynamics of the melt pool [6–10]. Modeling, monitoring and controlling the melt pool characteristics have been carried out in two major categories: *ex situ* modeling and *in situ* monitoring methods. *Ex situ* modeling is carried out after the part has been produced, while *in situ* monitoring takes place during the manufacturing process of the component [11]. In addition to the difficulties of building accurate *ex situ* models, parameters involved in a metal AM process are observed to have temporal fluctuations during fabrication [12–14]. For example, an observation revealed that the laser power (LP) used during the process can be approximately 20% less than the initially set value [15]. These fluctuations add uncertainty to the process that cannot be captured and modelled by *ex situ* modeling techniques. Whereas, with the development of sensing devices and control systems, *in situ* monitoring and controlling are employed to minimize the effect of modeling errors, as well as the uncertainties and fluctuations during the process [16,17].

Several research projects reported the use of *in situ* monitoring to control DED processes. Farshidianfar et al. [18] controlled the cooling rate of the melt pool using an infrared (IR) camera. They observed significantly less variation in the grain size and thereupon microhardness for parts produced under closed-loop conditions as opposed to that of the open-loop conditions. Smoqi et al. [19] controlled the melt pool temperature (MPT) using feedback signals from a coaxial two-wavelength imaging pyrometer. They have found that the parts built under closed-loop control have reduced variation in porosity and uniform microstructure compared to parts built under open-loop conditions. Although Fang et al. [20] investigated the response of the bead width in the CMT-based wire arc additive manufacturing (AM), specifically to understand how the bead width is sensitive to changes in process parameters, examining the dynamics of DED-LB/w to process parameter modification is yet to be performed.

It is essential to highlight that previous studies on system identification in metal AM have generally assumed a linear model when establishing the dynamic relationship between process parameters and signal outputs (process signatures) [21–23]. This study is to go beyond these assumptions to identify and select the best model that offers the highest level of accuracy.

By modeling and controlling the bead width, researchers have tried to control the bead thickness, minimize the surface waviness, and produce net shape parts [24]. To control the final geometry of parts, most of the researchers modelled and controlled the melt pool width (MPW) [25–27]. The MPW and bead width are strongly correlated; however, it is important to note that they are distinct entities. The MPW is a process signature that can be monitored and controlled while the material is being deposited whereas the bead width is the final quality of the part. The width of the melt pool and the bead can vary for several reasons, including shrinkage caused by temperature fluctuations and the presence of residual stresses within the part [28].

The primary objective of this study is to develop a modeling scheme suitable for closed-loop *in situ* control of the LM-DED process. This scheme will allow for the control and regulation of final properties that cannot be directly measured during the deposition process. Termed as the "parameter-signature-property modeling scheme," this approach establishes a clear connection between various elements. In this context, the term "parameter" refers to process parameters that can be adjusted and modified throughout the deposition process. On the other hand, "signature" pertains to material or process characteristics that are linked to the final properties and can be measured *in situ* during deposition. Examples of process signatures include the MPW, melt pool length (MPL), MPT, and acoustic emittance. Lastly, "properties" refer to the essential characteristics of the final product that require control, but unfortunately, cannot be measured during the deposition phase. Some examples of such properties include yield strength, residual stress, and final geometry.

It is important to mention that the parameter-signature-property relationship defined in this study is different from the process-structure-property (p-s-p) relationship that is used in the literature [29–31]. In the p-s-p relationship, the middle word "structure" refers to the part's structural properties, such as microstructure, phase configuration, and grain morphology. On the other hand, in the parameter-signature-property relationship, the linking bridge is the "process signature," a signal measured during deposition that exhibits a strong correlation with the final property. Another notable contrast is the purpose of the two modeling approaches. The p-s-p relationship is often employed for modeling and simulating the manufacturing process. In contrast, the paramount objective of defining the parameter-signature-property relationship is its practical application in the closed-loop control of final properties. By recognizing and comprehending these distinctions, this study's modeling scheme opens up new avenues for control and regulation of critical final properties during metal AM, leveraging the unique capabilities of process signatures to achieve the desired properties.

Figure 1 graphically illustrates the triple relationship in the parameter-signature-property modeling scheme. In this depiction, $F_1$ represents the function that defines the relationship between the process parameters and signatures. Similarly, $F_2$ denotes the function that connects process signatures to final properties. Throughout the deposition, the process signatures, which are essentially sensor signals, can be continuously monitored, providing valuable insights and strong correlations with the final properties. By establishing this comprehensive modeling scheme, the study aims to enable effective closed-loop control, allowing for precise regulation of critical final properties during the deposition process.

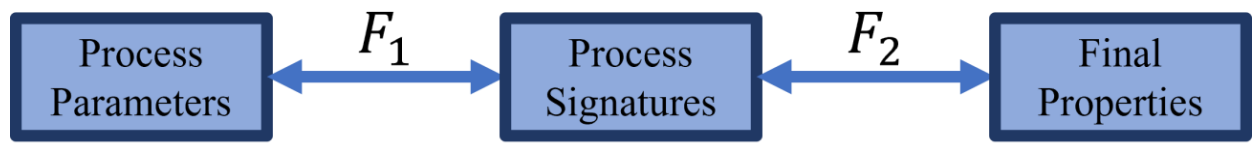

*Figure 1. Relationships between process parameters, process signatures, final properties, and the definition of $F_1$ and $F_2$.*

In this paper, the dynamic responses of melt pool characteristics on different process parameters are investigated and different linear and nonlinear transfer functions are introduced and compared to select the most appropriate one for the DED-LB/w. Finally, using the parameter-signature-property relationship, proportional-integral-derivative (PID) controllers are designed to control the bead width of the parts and the applicability and effectiveness of the proposed method is tested. The objectives of this study are defined as below:

- To investigate and compare the dynamic response of the MPW, MPL, and MPT to the change in different process parameters,
- To assess different linear and nonlinear dynamic models between process parameters and signatures ($F_1$),
- To develop, compare, and validate different modeling schemes between process signatures and the final bead width ($F_2$),
- To integrate $F_1$ and $F_2$ in the closed-loop control and designing a controller to control the bead width of the part, and
- To verify the effectiveness and usefulness of the proposed parameter-signature-property relationship in process control.

The remainder of the paper is organized as follows. Section 2 explains the methods and materials used in this research. Section 3 presents the results of the experiments and models and discusses the findings and observations. Finally, conclusions are drawn in Section 4.

## 2. Materials and Methods

### 2.1. Experimental setup and material

To examine the system's dynamic response to variations in different process parameters, four single and four multi-layer beads are printed. Experiments are carried out in a cell that consists of a six-axis IRB-4400 robot (ABB Robotics, Västerås, Sweden) and a laser optic cube that is installed on the robot's arm, wire feeder and an electrical power source system (EA Elektro-Automatik GmbH & Co KG, Viersen, Germany) that preheats the wire. Figure 2 illustrates the in-house robotized DED-LB/w cell. The movement of the wire is controlled using a wire feeder T drive 4 Rob 3 (EWM AG in Mündersbach, Germany) along with an AM8131 servo-drive (Beckhoff in Verl, Germany). The CX9020 PLC (Beckhoff, Verl, Germany) is utilized for process control and actuator communication.

For the laser source, a 12 kW TruDisk fiber laser (Trumpf, Ditzingen, Germany) is employed in continuous wave mode, emitting laser light at a wavelength of 1070 nm and the nominal diameter of 5 mm in focus. The incident angle of the beam is adjusted away from the normal direction to the workpiece surface to prevent back-reflections into the focusing optics. A CMOS camera (Photonfocus), that is coaxially inserted in the laser optics, receives the light emitted after it is reflected by a reflective golden mirror. The CMOS camera is used to capture the melt pool during the process. The temperature of the deposited metal during deposition is measured using a digital 2-color IMPAC pyrometer with fiber optic that is set in place on the robot arm. The IMPAC pyrometer is a non-contact measuring tool with a laser aiming light that continually points at the deposited metal at 20 millimeters before the laser spot. LabView software is used to extract temperature data as well as pictures from the camera.

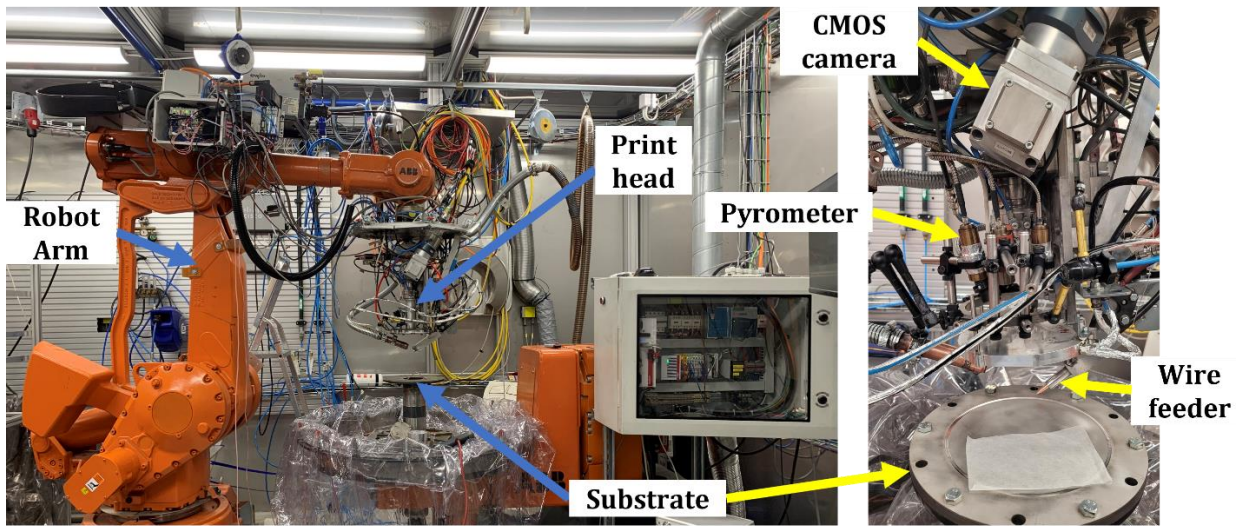

*Figure 2. DED-LB/w experimental setup with an overall view (left) and a detailed view of the head (right).*

Throughout the deposition process, an adaptive control system is employed to maintain a constant conductance (G) between the filler wire contact tube on the processing tool and the melt pool for process stability [32,33]. Achieving this stability involves adjusting two key parameters: the distance between the wire and the printing bed, and the wire feed speed (WFS). The automatic control algorithm utilized for this purpose is implemented by PROCADA AB, a company based in Trollhättan, Sweden. Furthermore, the wire is pre-heated using resistive heating. It has been demonstrated that utilizing two separate heat sources, namely the laser beam and electrical power (EP) for resistive pre-heating, enables precise control of the heat input, enhances the depth of penetration, and mitigates the occurrence of lack of fusion defects [34].

## 2.2. Parameters and geometry

This study investigates the dynamic response of the melt pool to variations in four distinct process parameters, including WFS, LP, EP, and travel speed (TS). To explore their impact, four beads and four five-layer walls are printed with the length of 120 mm. Within each bead, one of the process parameters is altered twice, while the remaining parameters are held constant. The parts are printed such that the travel direction is in the X axis and the build direction is in the Z axis. The deposited material is duplex stainless steel 2209 (DSS2209) with the wire diameter of 1.2 mm and the substrate material is stainless steel 316L [35].

In Table 1, each process parameter and its corresponding value are presented for each segment of the print. The range of process parameters are selected in a way that the stability of the process is maintained while the maximum change in the response (MPW and bead width) can be seen during the deposition of each part. Within each bead, one specific process parameter transitions from value A to B, and then from value B to C. These transitions occur at intervals of 40 mm, indicating that each segment has a length of 40 mm. The process parameter that changes within each bead and its corresponding value for each segment are indicated in bold in Table 1. Following the completion of the single bead prints, four five-layer walls are subsequently printed using similar process parameters as those employed for the previously printed beads. To be specific, for each set of process parameters corresponding to a specific "Bead number" in Table 1, both a single bead and a five-layer wall are printed. It is worth noting that the step height for all the walls is set to be 0.9 mm.

*Table 1. Process parameters and their values (boldfaced rows show changing values).*

| Bead number | Process parameter | Segment | | |
|---|---|---|---|---|
| | | A | B | C |
| 1 | LP (W) | 3000 | 3000 | 3000 |
| | WFS (m/min) | 2 | 2 | 2 |
| | EP (W) | 100 | 100 | 100 |
| | **TS (mm/s)** | **10** | **12** | **8** |
| 2 | LP (W) | 3000 | 3000 | 3000 |
| | WFS (m/min) | 2 | 2 | 2 |
| | **EP (W)** | **50** | **150** | **50** |
| | TS (mm/s) | 10 | 10 | 10 |
| 3 | LP (W) | 3000 | 3000 | 3000 |
| | **WFS (m/min)** | **1.8** | **2.2** | **1.8** |
| | EP (W) | 100 | 100 | 100 |
| | TS (mm/s) | 10 | 10 | 10 |
| 4 | **LP (W)** | **2800** | **3200** | **2800** |
| | WFS (m/min) | 2 | 2 | 2 |
| | EP (W) | 100 | 100 | 100 |
| | TS (mm/s) | 10 | 10 | 10 |

## 2.3. Image processing and signal acquisition

The images captured by the CMOS camera are imported into Python 3 for the purpose of determining the MPW and MPL in each frame. To isolate the region containing the melt pool, each image is initially cropped. Subsequently, the cropped images undergo a binarization process utilizing the mean intensity of the pixels as the threshold value. The resulting image is considered as the melt pool. The MPW and MPL are then calculated as the diameters of the largest inscribing and the smallest enclosing circles of the melt pool, respectively. For a visual representation of these procedures, refer to Figure 3(a). Additionally, the pyrometer's output signal is scaled to $500^o - 2500^o C$ and is referred to as the MPT throughout this paper.

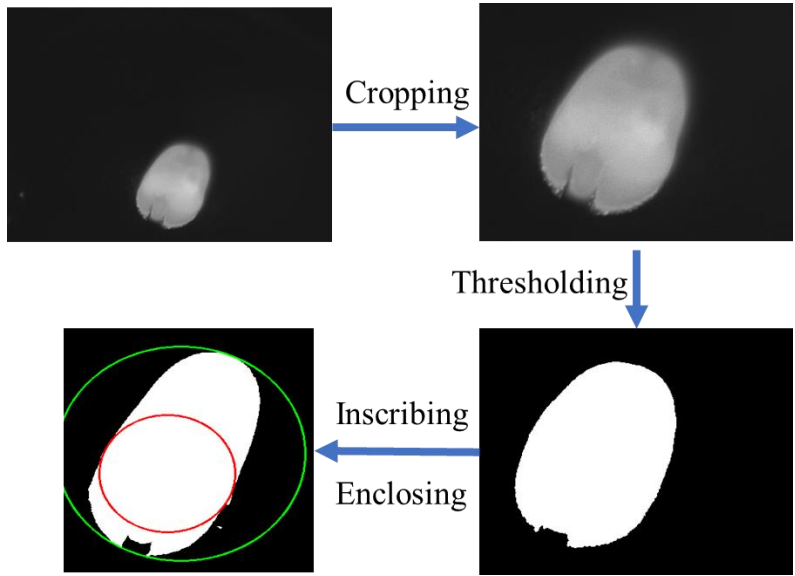

*Figure 3. Image processing procedure to find MPW and MPL.*

## 2.4. DED-LB/w system identification ($F_1$ construction)

System identification involves constructing mathematical models of dynamic systems based on measurements obtained from the input and output signals of a dynamic system [36]. In this study, system identification is utilized to discover and compare the dynamic model connecting process parameters and signatures ($F_1$). With reference to the parameter-signature-property relationship described in the Introduction, the resulting dynamic models, known as $F_1$, can subsequently serve as a foundation for the controller design and simulation of the DED-LB/w system.

As the purpose of this study is to control and model the width of the part, the MPW is selected as the process signature to be modeled. The synchronized MPW and process parameters are input to the System Identification MATLAB toolbox to examine different linear and nonlinear dynamic models including first and second order linear models, Hammerstein-Wiener models, and ARX (autoregressive with extra input) models with different mapping functions. The transfer function of the first order models with a delay term is explained by Equation 1,

$$G(s) = \frac{K}{1+T_w s} e^{-T_d s} \tag{1}$$

Where, $K$ is the steady state gain between the input and output. $T_W$ and $T_d$ are the time constant and delay time, respectively. The time constant is the duration in which the dynamic system approximately reaches 63.2% (i.e., $(1-e^{-1}) \times 100$ at one time constant $t = T_W$) of its steady-state response and it shows how quickly the system responds to the change in the input parameter.

The Hammerstein-Wiener structure is a configuration that comprises three blocks: a linear dynamic block sandwiched between two nonlinear steady-state blocks. These nonlinear blocks are referred to as the input nonlinear function and the output nonlinear function. In this structure, all three blocks collectively replace the traditional process block [37]. Moreover, ARX models characterize the observed system output as a mapping function—either linear or nonlinear—of regressors derived from past input and output observations [36,38]. The block diagram of both Hammerstein-Wiener and ARX models are shown in Figure 4. The best fit (BF) value, the R-squared value of the models expressed as a percentage, is used to compare the model accuracy.

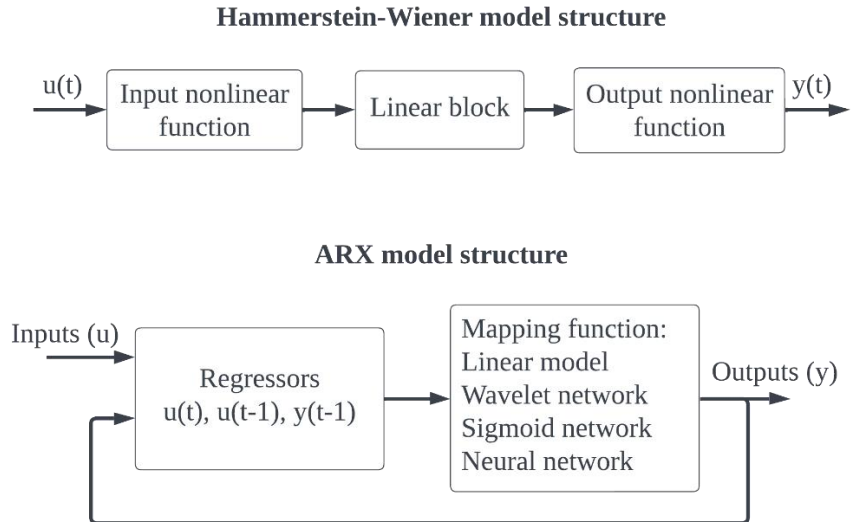

Figure 4. Hammerstein-Wiener and ARX model structures.

The system is treated as a single-input single-output (SISO) system since only one parameter is altered during the deposition of each part, which leads to the development of distinct models for each process petameter.

### 2.5. Bead width extraction and $F_2$ construction

To complete the parameter-signature-property relationship, the model between process signatures (MPW, MPL, and MPT) and final property (bead width) are constructed ($F_2$ model). In the following paragraphs, the steps to develop the model between melt pool signatures and bead width is described.

The printed beads and walls undergo scanning using a GOM ATOS Compact Scan 3D scanner. Subsequently, the resulting cloud points are imported into Autodesk Meshmixer to fill in unscanned portions and fix any possible issue. The point cloud data of the beads and walls is then imported into SOLIDWORKS.

The process of extracting the bead width at each point for each layer is illustrated in Figure 5. Initially, the walls are sectioned by employing cutting planes aligned parallel to the XY plane. The spacing between consecutive cutting planes equals to the step height of the print, which is 0.9 mm. Next, the intersection curve is obtained by intersecting the point cloud data with the cutting plane. A total of 1,000 evenly distributed points are extracted along each curve and exported to MATLAB utilizing a SOLIDWORKS macro script. For each layer, a spline is fitted to the points, enabling the calculation of the width at each point (x) based on the coordinate system of the points.

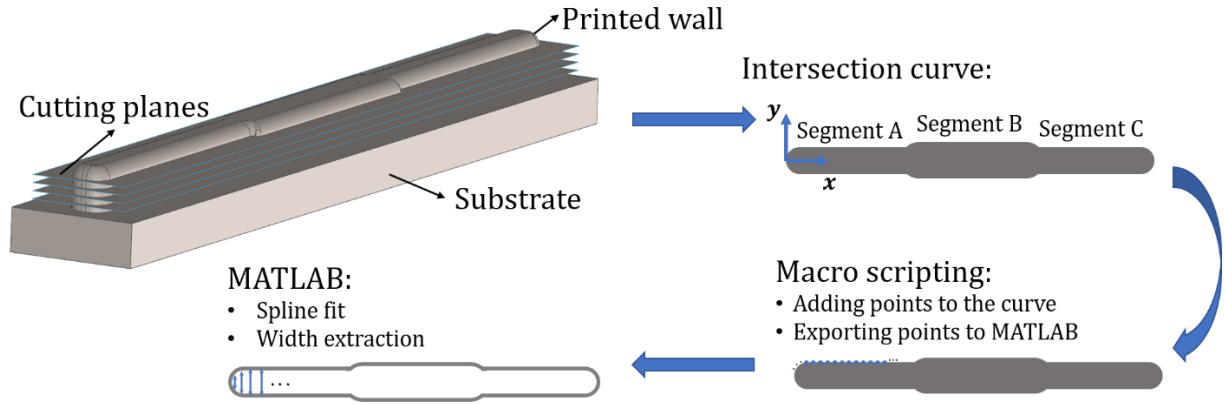

*Figure 5. Steps to extract the bead width at each point (x) of the walls.*

The next step is to synchronize the bead width with the signal data, specifically the melt pool signatures. The time series data is sampled at a rate of 0.03 seconds. To extract the bead width, it is ensured that the distance between two consecutive data points corresponds to the distance covered by the laser head in 0.03 seconds. The data from the printed walls is extracted and synced with the corresponding melt pool signatures.

The model structure utilized in this stage is a well-established multi-layer perceptron (MLP). MLP is a prominent variant of artificial neural networks, finding extensive application in regression modeling. Its utility extends to the field of predicting properties and structural attributes of parts produced through Metal AM [39]. The regression model to be developed is represented by Equation 2, where $BW$ denotes the final bead width at each time and $n$ indicates the layer number.

$$BW = F_2(MPW, MPL, MPT, n) \qquad (2)$$

$F_2(.)$ is trained using 80% of the available data, while the remaining data is allocated for validation purposes. Further details regarding the specifications of the neural network and comparison between different choices of input parameters for $F_2$ are discussed in Section 3.3.

### 2.6. Closed-loop control of the bead width

After developing the process transfer function ($F_1$) and the signature-property model ($F_2$), these models are incorporated into a control block diagram to design a controller for regulating the bead width of multi-layer walls. Simulation of this process is performed using

SIMULINK in MATLAB, considering two seconds to print each layer. The printing process initiates at t = 1 sec, resulting in a total simulation duration of 11 seconds for five layers.

Figure 6 (a) illustrates the control block diagram of the DED-LB/w process, where $F_1$ is utilized as the transfer function (parameter-signature link), and $F_2$ serves as the signature-property link. By integrating both models, the controller is expected to optimize the process parameters to achieve the desired bead width. On the other hand, Figure 6 (b) displays the control block diagram of the DED-LB/w, neglecting the signature-property relationship ($F_2$). Instead, it focuses on controlling the MPW to regulate the bead width. To compare the final bead width between the two scenarios, Figure 6 (b) includes the addition of $F_2$ after $F_1$. This enables the calculation of the bead width based on the output obtained from $F_1$. The calculated bead width can then be compared with the bead width obtained in the first scenario (Figure 6 (a)).

This comparative analysis allows for an evaluation of the influence of $F_2$ on the final bead width. It helps determine the effectiveness of the control strategy and assess whether the inclusion of $F_2$ enhances the control of the DED-LB/w.

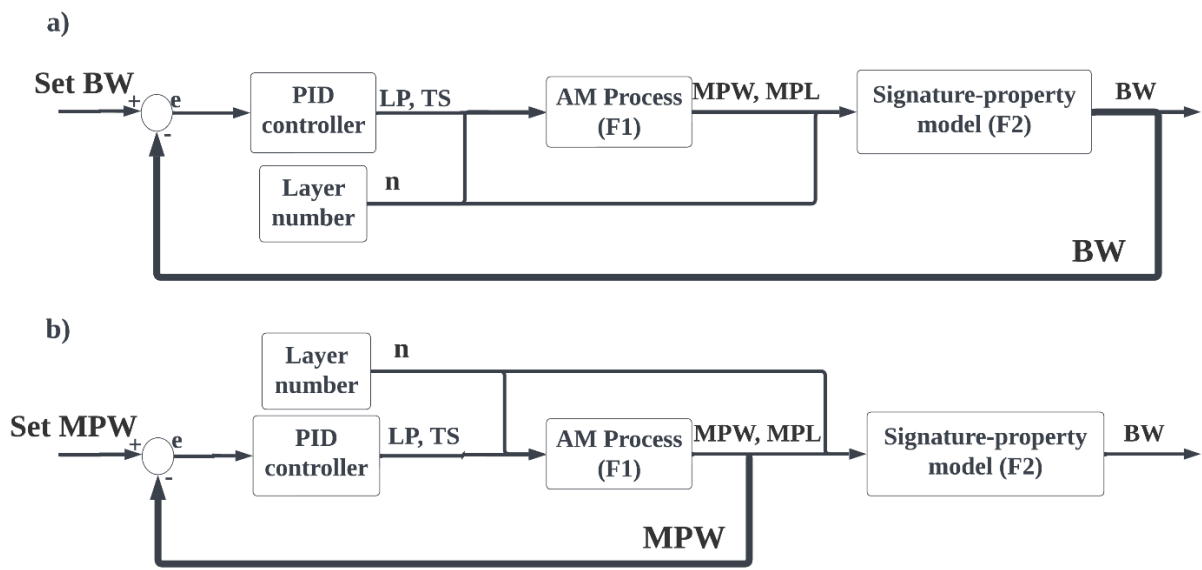

Figure 6. Control block diagram of the DED-LB/w. (a) Controlling the final bead width of the process using the parameter-signature-property relationship. (b) Controlling the MPW and ignoring the signature-property model ($F_2$).

PID controllers are utilized for controlling the width during the deposition. The closed-loop transfer function is linearized around the point where the controlled parameter (either the bead width or MPW) is five millimeters. The controller gains are then tuned to minimize overshoot and rise time in response to a unit step input. The relationship between the controller outputs (process parameters) and inputs (error) are explained in Equation 3,

$$u(t) = K_p e(t) + K_i \int e(t)dt + K_d \frac{de(t)}{dt} \qquad (3)$$

Where $K_p$, $K_i$, and $K_d$ are proportional, integral, and derivative gains, respectively. Moreover, $u(.)$ is the manipulated process parameter and $e(.)$ is the error between the controlled parameter set and real time value. The effectiveness of the proposed parameter-signature-property method is then evaluated by comparing the controlled bead width calculated from both of the scenarios with the desired bead width.

## 3. Results and Discussion

Four beads and four five-layer walls of 120 mm in length are printed using the established DED-LB/w setup. The visual appearance of the beads and walls from the top view and the number of each part are shown in Figure 7. The points at which the process parameters are altered are shown by dashed lines. Moreover, the process parameter that is altered during the deposition of each part is shown on the right side of Figure 7.

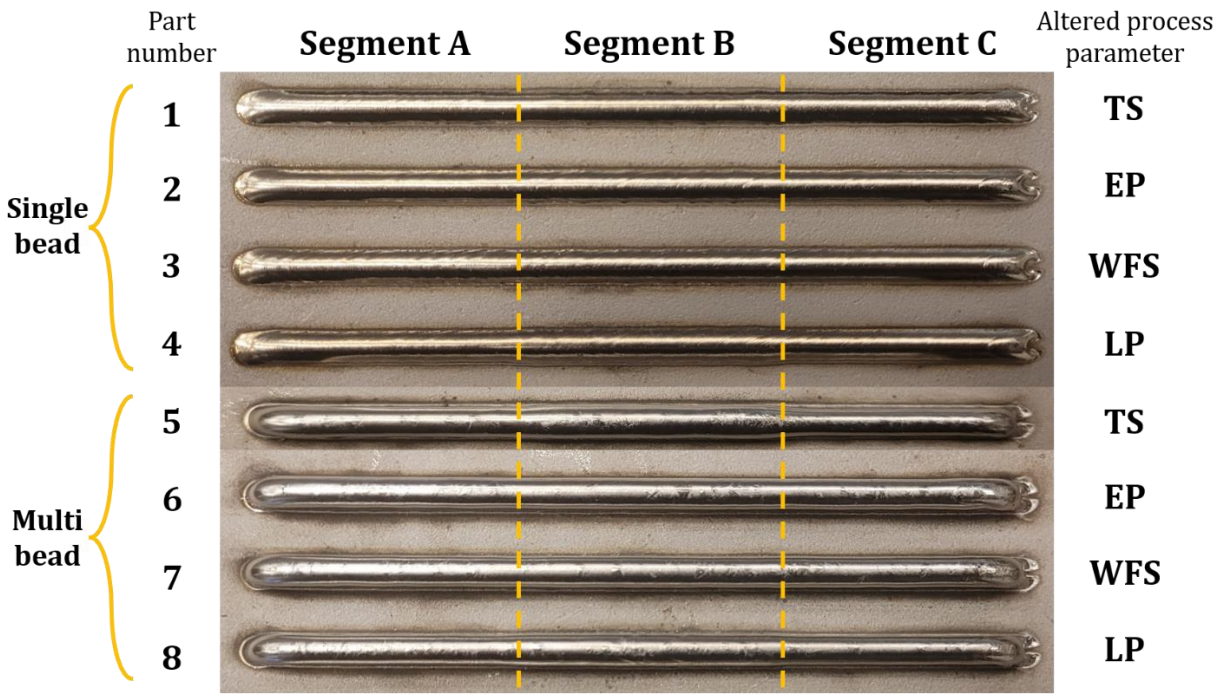

*Figure 7. Top view of the printed beads and walls. The process parameter that is altered during the deposition of each part is shown on the right.*

Sections 3.1 and 3.2 delve into the dynamic response of melt pool characteristics when different process parameters are modified, specifically focusing on single-bead analyses (parts 1-4) and multi-layer wall examinations (parts 5-8), respectively. Section 3.3 describes the construction of the signature-property model ($F_2$), and Section 3.4 focuses on the closed-loop control of the bead width.

## 3.1. System identification and analysis of single beads

The melt pool images are processed and the MPW and MPL at each time are calculated for all the printed single beads. The MPW during the deposition of each of the four single beads is presented in Figure 8. All plots depicted in Figure 8 are subjected to noise reduction through filtering by a simple moving average filter, meaning that each data point is replaced by the average value of the surrounding eight data points, and both the filtered signal and the original data are plotted. Findings depicted in Figure 8 indicate that altering the EP and WFS does not result in any significant change in the MPW. However, variations in the MPW are observed when alterations are made to the TS and LP.

The clear influence of LP and TS on the bead width is anticipated as observed in previous research studies [40,41]. Similar studies also indicated a positive-monotonic relationship between WFS and bead width [41,42]. However, upon examining Figure 8, it becomes evident that modifying the WFS does not result in noticeable changes in the bead width. This unexpected finding can be attributed to the implementation of an automatic control system in the DED-LB/w system. The control system adjusts the WFS to regulate the conductance (G) between the wire head and the melt pool. Consequently, altering the WFS set value during the deposition of different segments does not lead to significant variations in the bead width. Regarding the EP, while previous studies have indicated that pre-heating the wire impacts bead dilution and width-to-height ratio, our findings align with other research in suggesting that pre-heating the wire does not significantly affect the bead width [34,43]. Considering the finding that EP and WFS do not significantly affect the melt pool, and consequently, the bead width, the subsequent focus of this study centers around modeling and analyzing data obtained from the parts that are printed with variations in TS and LP, namely, parts 1, 4, 5, and 8.

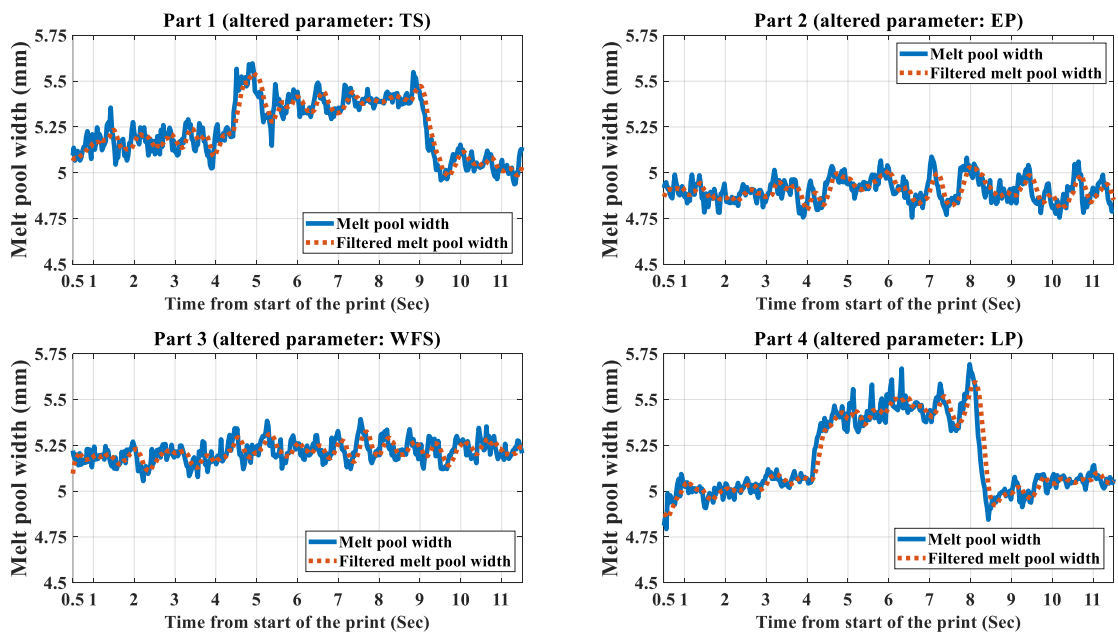

*Figure 8. Influences of process parameters on the MPW for single bead parts (Parts 1-4).*

Figure 9 displays the MPW and MPL of Parts 1 and 4, representing beads in which TS and LP are respectively altered. The left plots in Figure 9 present the MPW and MPL of Part 1, while the right plots exhibit the MPW and MPL of Part 4. To compare the dynamic system responses to changes in process parameters, the modified process parameter during the deposition of each bead (i.e., TS or LP) is included in the plots. It is important to mention that since Parts 1 and 4 are single beads and the substrate is cold prior to deposition, the temperature recorded by the pyrometer does not reach a minimum detectable temperature of 500°C. Consequently, the pyrometer could not provide meaningful information, and therefore, the temperature data is not included in Figure 9.

The first row of plots in Figure 9 portrays the MPW during the deposition process. Although changing both TS and LP lead to the change in the response, the MPW is more responsive to changes in LP, as indicated by the significantly shorter delay between the dashed orange line and the MPW in the right plot compared to the delay observed between the MPW and the TS command in the left plot. Furthermore, the rise time of the system in Part 4 is notably shorter than that of Part 1. This implies that LP can effectively control the MPW not only with less delay but also more rapidly compared to the TS.

The second row of plots in Figure 9 illustrates the MPL during the deposition of Parts 1 and 4. Increasing the LP from 2800 W to 3200 W results in an increase in the quasi-steady state response of the MPL from 9.5 mm to 10.4 mm, indicating an approximately 9.5% change in the length of the melt pool. From the MPL of Part 1 (the left down plot in Figure 9), it can be inferred that although the quasi-steady state response of MPL changes by changing TS, the amount of change is not significant. Hence, the TS does not exert control or regulation over the MPL.

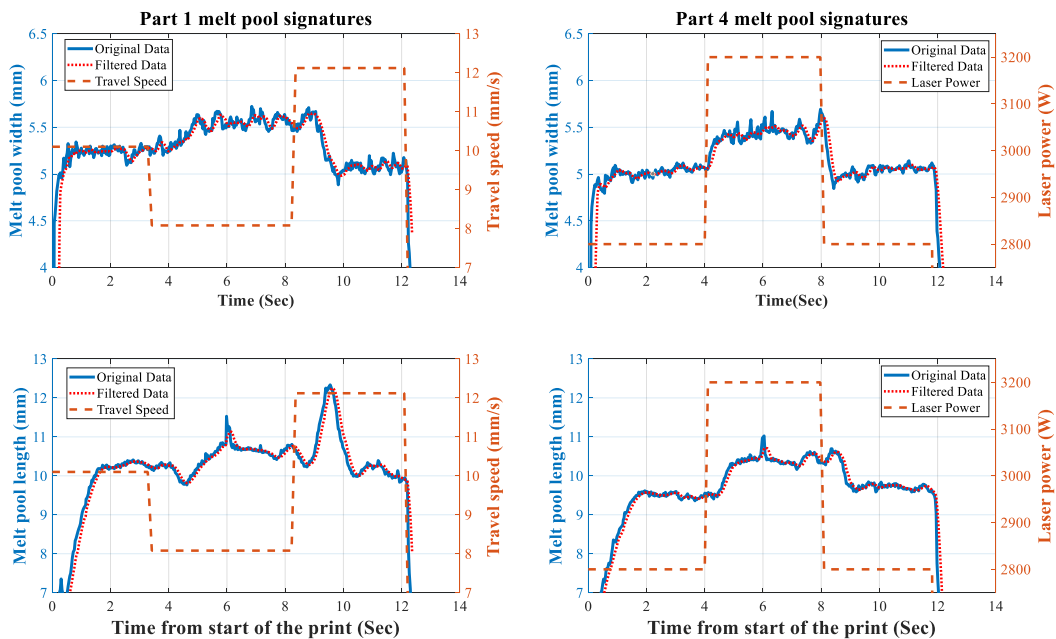

*Figure 9. Melt pool widths and lengths for Parts 1 (left) and 4 (right) with changes of TS and LP.*

After analyzing the behavior of the MPW and MPL, system identification is performed for single beads, assuming a single-input single-output system. To compare and validate different models, half of the data of each bead is utilized as the validation data.

Figure 10 displays the predictions of first and second order linear dynamic models between the MPW and process parameters while depositing the single beads. On the left plot, the dynamic models between the MPW and TS are compared, while the right plot compares the models between the MPW and LP. To enhance model accuracy, the data is scaled, and the mean is subtracted. As a result, the Y axes in Figure 10 do not represent the MPW in millimeters. During the modeling procedure it is found that nonlinear ARX and Hammerstein-Wiener models do not generate models with good accuracies and hence are not included in Figure 10.

The first order model outperforms the second order linear model, as indicated by the higher BF values. It appears that the behavior of the MPW in single layer beads can be accurately represented by a first-order linear model. The first order transfer function between the process parameter (LP and TS) and MPW is shown on both plots in Figure 10. The comparison of the time constants and delay times in the transfer functions show that the LP is a more suitable parameter to control the MPW.

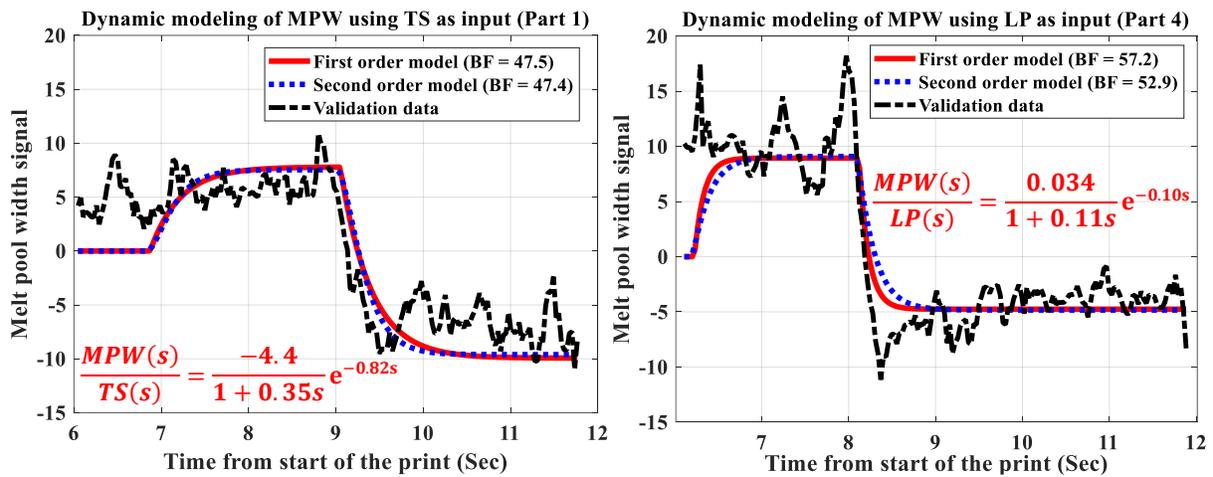

Figure 10. Dynamic linear model predictions of MPW for single bead printing using TS (left) and LP (right) as input parameters.

## 3.2 System identification and analysis of multi-layer walls

Four five-layer walls are printed, and one process parameter is altered during the printing of each wall. Figure 11 shows the width, length, and temperature of the melt pool while Parts 5 and 8 are deposited. As a reminder, Parts 5 and 8 are the multi-layer walls that have the altered TS and LP during their deposition, respectively. The melt pool signatures of Part 5 are depicted on left plots while the melt pool signatures of the Part 8 are on the right. On each plot of Figure 11, the solid blue line is a melt pool signature, and the dashed orange line is the process parameter whose axis is on the right side of the plot. On each plot the deposition of five layers can be seen.

It can be seen from the first row of the plots in Figure 11 that by increasing the layer number the MPW decreases. When the first bead starts to deposit on the baseplate, the melt pool has more freedom to move on the flat surface of the baseplate hence the width of the melt pool is higher in the first layers. As the layer number increases, the melt pool has less available flat surface to spread through and hence the melt pool width decreases by increasing the layer number. This observation aligns with the behavior of the walls printed in other studies [25,44].

The plots in Figure 11's second row illustrate the MPL during the deposition of Parts 5 and 8. It is observed that as the layer number increases, the MPL increases as well, as indicated by both plots. Moving to the third row of plots in Figure 11, it is noteworthy that the behavior of the MPL and MPT exhibits similarities, where an increase in MPT results in a longer melt pool. This observation aligns with the study performed by [45] since they found a linear relationship between the melt pool area and MPT. This finding has practical implications as it enables researchers to minimize the amount of sensory equipment required, as they can rely on the MPL to estimate the temperature.

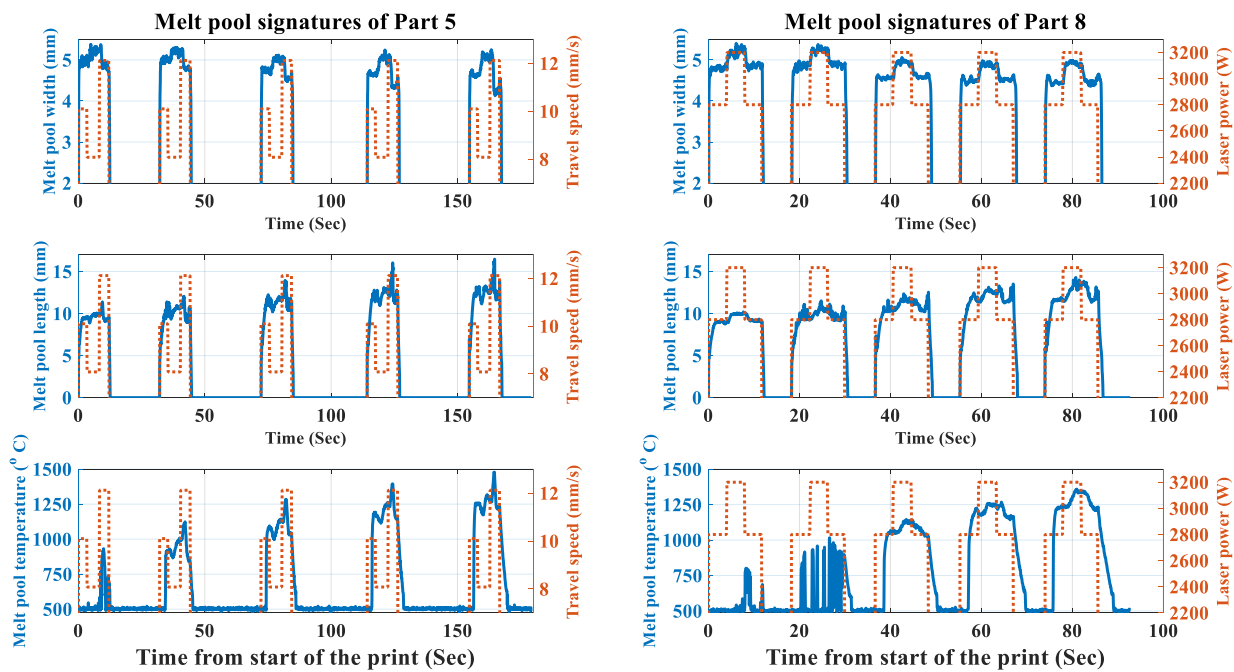

*Figure 11. MPW, MPL, and MPT of the walls in which the TS (left) or LP (right) is changed.*

The next step is to develop the dynamic models between the MPW and process parameters ($F_1$) in multi-layer walls. The signal acquisition and modeling are only performed for the wall in which the LP is manipulated (Part 8) as the LP is found to be the most impactful process parameter. The models developed in this stage, unlike those in Section 3.1, are designed to be applicable to multi-layer walls. These models are expected not only to enable the design of controllers for the DED-LB/w but also provide insights into its dynamics.

The MPW data of different layers in Part 8 is concatenated together to omit the zero data. Thereafter, it is filtered using a low-pass band filter with the cut of frequency of 13 Hz and the mean value of the signals are removed to increase the models' accuracy. The frequency is found by trial-and-error in a way that the trend of the data is preserved and the noise is also reduced. The continuous signals of the LP (input) and the MPW (output) are shown in the left plots of Figure 12. It can be seen that the LP only switches between 2800 W and 3200 W and it does not set back to zero. Moreover, about the last 30% of the data is used as the validation data which is the signal from the fourth and fifth layer and can be seen in Figure 12.

Modeling without incorporating the layer number as one of the input parameters does not lead to accurate model since the influence of the layer number on the behavior of the MPW was discovered to be significant. As can be seen from the figure, the moments at which the layer is raised, MPW signal shows a sudden drop. As a result, the layer number is included as an input parameter in the modeling scheme. Two transfer functions are developed simultaneously during the modeling process: one between the LP and MPW, and the other between the layer number and MPW. By incorporating both transfer functions, the dynamic behavior of the MPW is determined. This incorporation is described by Equation 4,

$$MPW(s) = G_{LP}(s)LP(s) + G_n(s)\frac{n}{s} \quad (4)$$

where, $MPW(s)$ and $LP(s)$ are the Laplace transform of the MPW and LP signals, respectively, and $G_{LP}(s)$ and $G_n(s)$ are the transfer functions between the LP and layer number as inputs and the MPW as the output. Moreover, $n$ is the layer number and $\frac{n}{s}$ is the Laplace transform of the layer number. After performing the modeling, it is found that using the first order linear transfer function with a delay as $G_{LP}(s)$ and a static gain as the $G_n(s)$ leads to the highest accuracy on the validation data. In the right plot of Figure 12, it can be seen that the red dotted line outperforms other models. The derived transfer functions of the best model (first-order linear model for LP and static gain for layer number) can be seen in Figure 12. As there is a reverse relationship between the layer number and MPW, a negative number is expected for the static gain, which is -0.11 in this case.

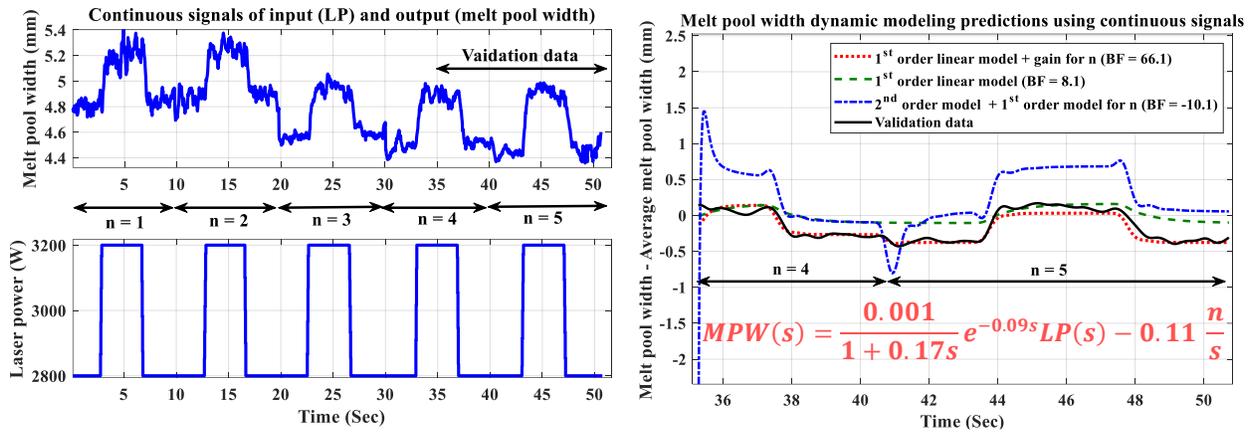

*Figure 12. The MPW and LP continuous signals (left) and the MPW dynamic model prediction on the validation data using the continuous signals (right).*

Thus far, the dynamic modeling of the relationship between process parameters and melt pool signatures ($F_1$) has been conducted. The subsequent section aims to model the relationship between melt pool signatures and the final property ($F_2$).

### 3.3 Signature-property modeling ($F_2$ construction)

In the signature-property model (F2), the process signatures, MPW, MPL, and MPT are the inputs and the bead width is the output, representing the product property. Bead width data for Parts 5 and 8 is extracted using the procedure outlined in Figure 5. The bead width data is then synchronized with the time series data of the melt pool characteristics, with a time step of 0.03 seconds. After the data synchronization for Parts 5 and 8, in total 4,035 data points are extracted. The dataset used to construct $F_2$ can be downloaded [here](here).

To model the bead width, Equation 2 is employed, utilizing four input parameters: the MPW, MPL, MPT, and layer number. Prior to modeling, the input data is normalized between zero and one. Moreover, to find the suitable transformation of the output data, different transformations are tested and finally the Napierian logarithm is selected. This transformation and normalization aim to enhance the accuracy of the final models. Additionally, all input data is filtered using a window size of eight.

As explained in Section 2.5, MLP or fully-connected neural network is utilized as the model. Different numbers of hidden layers and neurons are used and the MLP with the least root mean square error (RMSE), least mean absolute error (MAE), and highest R-squared ($R^2$) on the validation data is selected as the final model. Different numbers of hidden layers and neurons as well as different choices of activation functions for each layer are tested and, finally, the architecture shown in Figure 13 is selected. The final MLP consists of six hidden layers with 8, 16, 32, 16, 8, and 4 neurons, respectively. The activation function of the first hidden layer is set to be the rectified linear unit (ReLU) while other hidden layers are followed by a sigmoid activation function ($F(x) = \frac{1}{1+e^{-x}}$). The optimization method used to find the MLP weights is the Levenberg-Marquardt algorithm. This algorithm can be seen as a middle ground between the Gauss-Newton and gradient descent optimization methods and is widely used to train neural networks [46]. The training stopping criteria is set to reach 200 epochs.

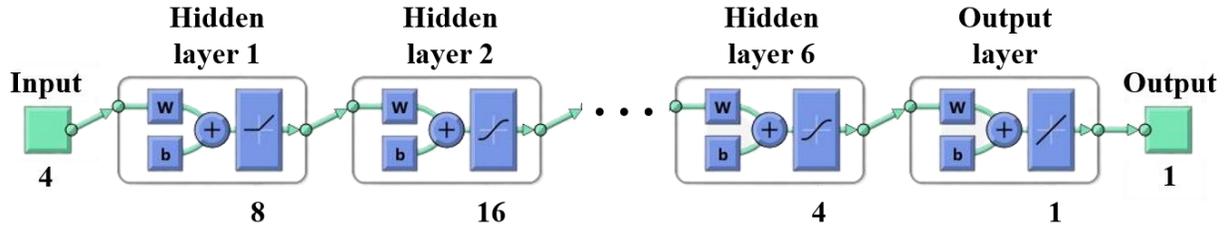

*Figure 13. Architecture of the MLP utilized to develop the $F_2$ model*

MLP is trained using 80% of the data and the remaining 20% is used to validate the final model. The final R-squared value of the MLP calculated using the validation data is 0.9853 with RMSE and MAE being 0.0256 and 0.0174 respectively, indicating a high accuracy of the developed model.

To justify the choice of MLP input parameters, two other neural networks are trained, one with the MPW and layer number as input parameters and one with MPW, MPL, and layer number. The RMSE, MAE, and R-squared values of these MLPs are compared with the original model in Table 2. It can be understood from Table 2 that using all the three melt pool signatures namely, the MPW, MPL, and MPT along with the layer number yields the model with the highest accuracy from all three metrics. This observation confirms the hypothesis that although the bead width of the single-track structures is shaped by the MPW during the deposition, other process signatures such as MPL, MPT, and the layer number have significant effect on the width of the part.

It is worth mentioning that the accuracy of the developed signature-property models ($F_2$) with and without MPT are close to each other (the first two rows in Table 2). This finding is due to the similar behavior of the MPL and MPT that is also observed in Figure 11. Based on this similarity, it can be inferred that including either the MPT or MPL as inputs to the model is sufficient to achieve accurate results. Adding both parameters does not lead to a significant improvement in the model's accuracy.

*Table 2. Comparison of RMSE, MAE, and R-squared values on validation data points with different selections of input parameters.*

| Model | RMSE (mm) | MAE (mm) | R-squared |
|---|---|---|---|
| $BW = F_2(MPW, MPL, MPT, n)^*$ | 0.0256 | 0.0174 | 0.9853 |
| $BW = F_2(MPW, MPL, n)$ | 0.0425 | 0.0330 | 0.9351 |
| $BW = F_2(MPW, n)$ | 0.1452 | 0.1258 | 0.5624 |

*$BW$ and $n$ are the bead width and layer number, respectively.

Another important question yet to be addressed is the effectiveness of the proposed parameter-signature-property modeling. One may consider using the process parameters as inputs and the final bead width as the output, namely, $F_3$. Such a $F_3$ model is developed between the TS, LP, and layer number as inputs and the bead width as the objective. The RMSE, MAE, and $R^2$ values of the $F_3$ model are shown in Table 3. To make a fair comparison, the accuracy of the $F_3$ should be compared with a model comprised of $F_1$ and $F_2$ which is $F_2(F_1(TS, LP, n))$. The latter is the model that is derived from the parameter-signature-property modeling. The comparison is made in Table 3. The accuracy of the $F_3$ model ($R^2$ = 0.3258) is comparably less than the model developed by the parameter-signature-property model ($R^2$ = 0.6534). This shows that selecting an informative process signature that has a strong correlation with the final property and can be measured accurately during the process helps significantly with the accuracy of the to-be-developed models.

Table 3. Comparison between $F_3$ and $F_2(F_1)$.

| Model | RMSE (mm) | MAE (mm) | R-squared |
|---|---|---|---|
| $BW = F_3(TS, LP, n)$* | 0.2975 | 0.2546 | 0.3258 |
| $BW = F_2(F_1(TS, LP, n))$ | 0.1235 | 0.1168 | 0.6534 |

*$BW$ and $n$ are the bead width and layer number, respectively.

The comparisons between different modeling schemes of $F_2$ and $F_3$ highlight two important points:

1. Attempting to control and model the DED-LB/w without *in situ* monitoring (i.e., using $F_3$) appears to be inadequate due to the process's complexity and uncertainties. *In situ* monitoring enables the capture and control of these uncertainties.
2. Employing process signatures as a connecting link between process parameters and the final properties (parameter-signature-property modeling) enhances process modeling accuracy and efficiency. Prior to modeling, it is vital to carefully select and utilize the most informative signatures that exhibit the highest correlation with the final property.

## 3.4. Closed-loop control implementation and comparison

Referring to each of the closed-loop control scenarios shown in Figure 6, the manipulated parameter is selected to be the LP and the controlled parameter is the bead width in scenario 1 and the MPW in scenario 2 where the signature-property link is ignored. We'd compare the two control scenarios and highlight the benefits of having the signature-property link ($F_2$) in the control.

The process is simulated for 11 seconds as described in Section 2.6. During the first five seconds of the deposition (t = 1-6 sec), the width set value for the controlled parameter is 5 mm. In the second five seconds of the deposition (t = 6-11 sec), the set value is reduced to 4.7 mm. This variation in the set value is intended to assess the adaptability and effectiveness of the controller.

When selecting $F_2$ to be utilized in the closed-loop control, the accuracies of the $F_2$ model with and without the MPT are close and adding the MPT as one of the $F_2$ inputs does not significantly increase the accuracy of the model. As a result, the $F_2$ model considering the MPW, MPL, and layer number ($F_2(MPW, MPL, n)$) is employed in the closed-loop control simulations.

It is important to note that using the previously developed MLPs would yield highly nonlinear closed-loop transfer functions, making the tuning and optimization of gains impractical. Instead of neural networks, a response surface model (RSM) is utilized to construct $F_2(MPW, MPL, n)$ and reduce the nonlinearity of the closed-loop control transfer function. The RSM employed is a 3rd-order polynomial comprising all terms. The accuracy of the developed model is quantified as follows: RMSE = 0.0634, MAE = 0.0435 mm, and $R^2$ = 0.8942. Comparing the $R^2$ values of the MLP (0.9351) and the RSM (0.8942), it is evident that replacing the MLP with the RSM does not significantly decrease the accuracy.

The models are implemented in Simulink and the controller gains are tuned to minimize overshoot and rise time in response to a unit step input. In both scenarios, the manipulated variable is the LP. The PID controller gains for either of the scenarios are provided in Table 4.

*Table 4. Tuned controller gains for both of the scenarios.*

| Gain value | Scenario 1 | Scenario 2 |
|---|---|---|
| **Proportional gain ($K_p$)** | 985.20 | 1539.0 |
| **Integrative gain ($K_i$)** | 3996.6 | 5711.4 |
| **Derivative gain ($K_d$)** | 19.250 | 31.092 |

Figure 14 presents the results of the closed-loop control for both scenarios. The left side illustrates the control of the bead width (first scenario), while the right side displays the control of the MPW (second scenario). The first row of the plots in Figure 14 depicts the controlled parameter (solid line) and its set value (dashed line) during the simulation. It can be seen that the controller is able to regulate the controlled parameter to the set values of the controlled parameters in both scenarios as the layer number increases.

In the second row of the plots in Figure 14, the width of the part in both of the scenarios are compared with the desired bead width. The desired bead width is shown with a dashed line in both of the plots. It is desired to have five- and 4.7-millimeter beads in the first- and second-five seconds of the deposition, respectively. It is evident that in the second scenario where the process is controlled without integrating the signature-property link ($F_2$), although the controller is able to adjust the controlled parameter (MPW), the width of the part (bead width) is not equal to the desired value shown by the dashed line. This finding proves the applicability and usefulness of the proposed parameter-signature-property modeling to control the geometry of the printed parts.

The third and fourth rows of the plots in Figure 14 show the manipulated process parameter (LP) and the layer number during the simulation, respectively.

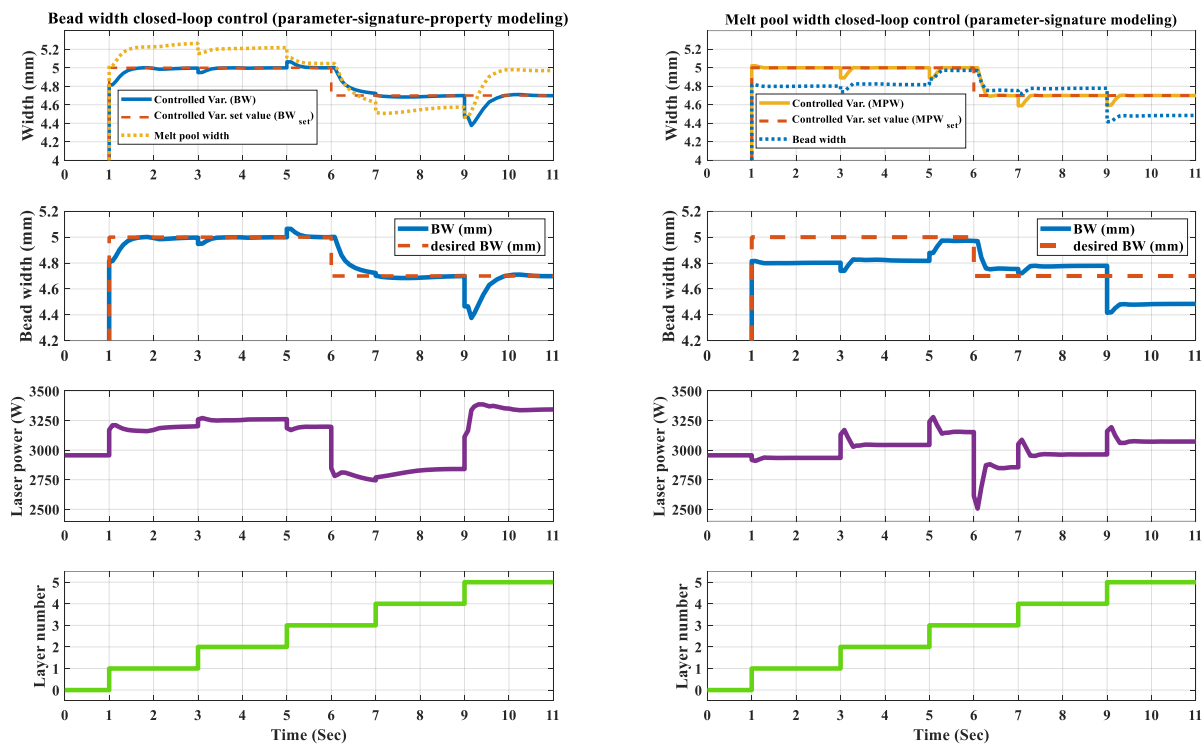

*Figure 14. Closed-loop control of bead width (left) and MPW (right).*

## 4. Conclusions

This study focuses on the dynamic modeling and closed-loop control for a part final property that cannot be measured *in situ* with the hot-wire directed energy deposition using a laser beam (DED-LB/w) process. The research investigates four process parameters, travel speed (TS), electrical power (EP), wire feed speed (WFS), and laser power (LP), to control and model the melt pool width (MPW) and, ultimately, the bead width of single- and multi-layer beads.

Different from other studies, this work proposes a parameter-signature-property relationship utilizing process signatures such as melt pool characteristics as connecting links between the process parameter and final properties that cannot be measured *in situ*. The effectiveness of the proposed parameter-signature-property modeling scheme is demonstrated by comparing the accuracy of the predicted final property (bead width) with and without the utilization of the parameter-signature-property modeling. Moreover, the developed models are integrated into a closed-loop control of the bead width. The proposed approach is able to control the width of the final part better than an alternative control scheme where the signature-property link is disregarded. The comparison between the two control scenarios demonstrates the superiority of the parameter-signature-property modeling scheme in controlling the part's final properties. The proposed method shows its promises in controlling additive manufacturing machines to achieve the desired part property that cannot be directly measured *in situ* during the printing.

In future studies, experiments will be performed to develop multi-input multi-output (MIMO) dynamic models between the process parameters and signatures and to increase the accuracy and generalizability of the dynamic models developed in this study. This work is to be extended to model process signatures with more complex part properties such as strength, porosity, roughness, and so on.

## Acknowledgement

Authors would like to express their sincere gratitude to Petter Hagqvist and Almir Heralic from Procada AB, whose expertise and assistance were instrumental in the successful execution of the experiments and the analysis of the obtained findings. The authors gratefully acknowledge funding from the Natural Sciences and Engineering Research Council (NSERC) of Canada [Grant numbers: RGPIN-2019-06601] and Business Finland under Project #: 4819/31/2021 with affiliation to the Eureka! SMART project (S0410) titled "TANDEM: Tools for Adaptive and Intelligent Control of Discrete Manufacturing Processes."

## Author contributions

Mostafa Rahmani Dehaghani: Conceptualization, Methodology, Investigation, Programing, Writing-Original Draft, Validation, and Writing-Review and Editing

Atieh Sahraeidolatkhaneh, and Morgan Nilsen: Conceptualization, Methodology, Visualization, and Writing-Review and Editing

Yifan Tang and Pouyan Sajadi: Data Curation, Visualization, Formal Analysis, and Writing-Review and Editing

Fredrik Sikström and G. Gary Wang: Conceptualization, Supervision, Project Administration, Funding Acquisition, and Writing-Review and Editing

The first draft of the manuscript was written by Mostafa Rahmani Dehaghani and all authors commented on previous versions of the manuscript. All authors read and approved the final manuscript.

## Declaration of Generative AI and AI-assisted technologies in the writing process

During the preparation of this work, the authors used GPT-3.5, developed by OpenAI, in order to improve the readability and language of the article. After using this tool, the authors reviewed and edited the content as needed and take full responsibility for the content of the publication.

## References

[1]  Heralic A. Monitoring and Control of Robotized Laser Metal-wire Deposition. Chalmers Tekniska Hogskola (Sweden) PP - Sweden, 2012.

[2]  Medrano A, Folkes J, Segal J, Pashby I. Fibre laser metal deposition with wire: parameters study and temperature monitoring system. Proc.SPIE, vol. 7131, 2009, p. 713122. https://doi.org/10.1117/12.816831.

[3]  Froend M, Ventzke V, Riekehr S, Kashaev N, Klusemann B, Enz J. Microstructure and hardness evolution of laser metal deposited AA5087 wall-structures. Procedia CIRP 2018;74:131–5. https://doi.org/10.1016/j.procir.2018.08.062.

[4]  Garmendia I, Flores J, Madarieta M, Lamikiz A, Uriarte LG, Soriano C. Geometrical control of DED processes based on 3D scanning applied to the manufacture of complex parts. Procedia CIRP 2020;94:425–9. https://doi.org/10.1016/j.procir.2020.09.158.

[5]  Mani M, Lane B, Donmez M, Feng S, Moylan S, Fesperman R. Measurement Science Needs for Real-time Control of Additive Manufacturing Powder Bed Fusion Processes 2015. https://doi.org/10.6028/NIST.IR.8036.

[6]  Yu G, Gu D, Dai D, Xia M, Ma C, Chang K. Influence of processing parameters on laser penetration depth and melting/re-melting densification during selective laser melting of aluminum alloy. Appl Phys A 2016;122:891. https://doi.org/10.1007/s00339-016-0428-6.

[7]  Criales LE, Arısoy YM, Lane B, Moylan S, Donmez A, Özel T. Laser powder bed fusion of nickel alloy 625: Experimental investigations of effects of process parameters on melt pool size and shape with spatter analysis. Int J Mach Tools Manuf 2017;121:22–36. https://doi.org/10.1016/j.ijmachtools.2017.03.004.

[8]  Dilip JJS, Zhang S, Teng C, Zeng K, Robinson C, Pal D, et al. Influence of processing parameters on the evolution of melt pool, porosity, and microstructures in Ti-6Al-4V

alloy parts fabricated by selective laser melting. Prog Addit Manuf 2017;2:157–67. https://doi.org/10.1007/s40964-017-0030-2.

[9] Keshavarzkermani A, Marzbanrad E, Esmaeilizadeh R, Mahmoodkhani Y, Ali U, Enrique PD, et al. An investigation into the effect of process parameters on melt pool geometry, cell spacing, and grain refinement during laser powder bed fusion. Opt Laser Technol 2019;116:83–91. https://doi.org/10.1016/j.optlastec.2019.03.012.

[10] Scime L, Beuth J. Using machine learning to identify in-situ melt pool signatures indicative of flaw formation in a laser powder bed fusion additive manufacturing process. Addit Manuf 2019;25:151–65. https://doi.org/10.1016/j.addma.2018.11.010.

[11] Bayat M, Dong W, Thorborg J, To AC, Hattel JH. A review of multi-scale and multi-physics simulations of metal additive manufacturing processes with focus on modeling strategies. Addit Manuf 2021;47:102278. https://doi.org/10.1016/j.addma.2021.102278.

[12] Moges T, Ameta G, Witherell P. A Review of Model Inaccuracy and Parameter Uncertainty in Laser Powder Bed Fusion Models and Simulations. J Manuf Sci Eng 2019;141. https://doi.org/10.1115/1.4042789.

[13] Olleak A, Xi Z. Calibration and Validation Framework for Selective Laser Melting Process Based on Multi-Fidelity Models and Limited Experiment Data. J Mech Des Trans ASME 2020;142. https://doi.org/10.1115/1.4045744.

[14] Rahmani Dehaghani M, Tang Y, Gary Wang G. Iterative Uncertainty Calibration for Modeling Metal Additive Manufacturing Processes Using Statistical Moment-Based Metric. J Mech Des 2022;145. https://doi.org/10.1115/1.4055149.

[15] King WE, Barth HD, Castillo VM, Gallegos GF, Gibbs JW, Hahn DE, et al. Observation of keyhole-mode laser melting in laser powder-bed fusion additive manufacturing. J Mater Process Technol 2014;214:2915–25. https://doi.org/10.1016/j.jmatprotec.2014.06.005.

[16] Ahn D-G. Directed Energy Deposition (DED) Process: State of the Art. Int J Precis Eng Manuf Technol 2021;8:703–42. https://doi.org/10.1007/s40684-020-00302-7.

[17] Cai Y, Xiong J, Chen H, Zhang G. A review of in-situ monitoring and process control system in metal-based laser additive manufacturing. J Manuf Syst 2023;70:309–26. https://doi.org/10.1016/j.jmsy.2023.07.018.

[18] Farshidianfar MH, Khajepour A, Gerlich A. Real-time control of microstructure in laser additive manufacturing. Int J Adv Manuf Technol 2016;82:1173–86. https://doi.org/10.1007/s00170-015-7423-5.

[19] Smoqi Z, Bevans BD, Gaikwad A, Craig J, Abul-Haj A, Roeder B, et al. Closed-loop control of meltpool temperature in directed energy deposition. Mater Des 2022;215:110508. https://doi.org/10.1016/j.matdes.2022.110508.

[20] Fang X, Ren C, Zhang L, Wang C, Huang K, Lu B. A model of bead size based on the dynamic response of CMT-based wire and arc additive manufacturing process

parameters. Rapid Prototyp J 2021;27:741–53. https://doi.org/10.1108/RPJ-03-2020-0051.

[21] Hofman JT, Pathiraj B, van Dijk J, de Lange DF, Meijer J. A camera based feedback control strategy for the laser cladding process. J Mater Process Technol 2012;212:2455–62. https://doi.org/10.1016/j.jmatprotec.2012.06.027.

[22] Moralejo S, Penaranda X, Nieto S, Barrios A, Arrizubieta I, Tabernero I, et al. A feedforward controller for tuning laser cladding melt pool geometry in real time. Int J Adv Manuf Technol 2017;89:821–31. https://doi.org/10.1007/s00170-016-9138-7.

[23] Liang Z, Liao Z, Zhang H, Li Z, Wang L, Chang B, et al. Improving process stability of electron beam directed energy deposition by closed-loop control of molten pool. Addit Manuf 2023;72:103638. https://doi.org/10.1016/j.addma.2023.103638.

[24] Saboori A, Gallo D, Biamino S, Fino P, Lombardi M. An Overview of Additive Manufacturing of Titanium Components by Directed Energy Deposition: Microstructure and Mechanical Properties. Appl Sci 2017;7. https://doi.org/10.3390/app7090883.

[25] Gibson BT, Bandari YK, Richardson BS, Henry WC, Vetland EJ, Sundermann TW, et al. Melt pool size control through multiple closed-loop modalities in laser-wire directed energy deposition of Ti-6Al-4V. Addit Manuf 2020;32:100993. https://doi.org/10.1016/j.addma.2019.100993.

[26] Kim MJ, Saldana C. Thin wall deposition of IN625 using directed energy deposition. J Manuf Process 2020;56:1366–73. https://doi.org/10.1016/j.jmapro.2020.04.032.

[27] Borovkov H, de la Yedra AG, Zurutuza X, Angulo X, Alvarez P, Pereira JC, et al. In-Line Height Measurement Technique for Directed Energy Deposition Processes. J Manuf Mater Process 2021;5. https://doi.org/10.3390/jmmp5030085.

[28] Vundru C, Singh R, Yan W, Karagadde S. A comprehensive analytical-computational model of laser directed energy deposition to predict deposition geometry and integrity for sustainable repair. Int J Mech Sci 2021;211:106790. https://doi.org/10.1016/j.ijmecsci.2021.106790.

[29] Lindgren L-E, Lundbäck A, Fisk M, Pederson R, Andersson J. Simulation of additive manufacturing using coupled constitutive and microstructure models. Addit Manuf 2016;12:144–58. https://doi.org/10.1016/j.addma.2016.05.005.

[30] Yan W, Lin S, Kafka OL, Lian Y, Yu C, Liu Z, et al. Data-driven multi-scale multi-physics models to derive process–structure–property relationships for additive manufacturing. Comput Mech 2018;61:521–41. https://doi.org/10.1007/s00466-018-1539-z.

[31] Kouraytem N, Li X, Tan W, Kappes B, Spear AD. Modeling process–structure–property relationships in metal additive manufacturing: a review on physics-driven versus data-driven approaches. J Phys Mater 2021;4:32002. https://doi.org/10.1088/2515-7639/abca7b.

[32] Heralić A, Christiansson A-K, Lennartson B. Height control of laser metal-wire

deposition based on iterative learning control and 3D scanning. Opt Lasers Eng 2012;50:1230–41. https://doi.org/10.1016/j.optlaseng.2012.03.016.

[33] Hagqvist P, Heralić A, Christiansson A-K, Lennartson B. Resistance measurements for control of laser metal wire deposition. Opt Lasers Eng 2014;54:62–7. https://doi.org/10.1016/j.optlaseng.2013.10.010.

[34] Kisielewicz A, Pandian KT, Sthen D, Hagqvist P, Bermejo MA V, Sikström F, et al. Hot-wire laser-directed energy deposition: Process characteristics and benefits of resistive pre-heating of the feedstock wire. Metals (Basel) 2021;11. https://doi.org/10.3390/met11040634.

[35] Zhang D, Liu A, Yin B, Wen P. Additive manufacturing of duplex stainless steels - A critical review. J Manuf Process 2022;73:496–517. https://doi.org/10.1016/j.jmapro.2021.11.036.

[36] Ljung L. System Identification: Theory for the User. Prentice Hall PTR; 1999.

[37] Ławryńczuk M. Nonlinear predictive control for Hammerstein–Wiener systems. ISA Trans 2015;55:49–62. https://doi.org/10.1016/j.isatra.2014.09.018.

[38] Guidorzi R. Multivariable System Identification. From Observations to Models. Bononia University Press; 2003.

[39] Ackermann M, Haase C. Machine learning-based identification of interpretable process-structure linkages in metal additive manufacturing. Addit Manuf 2023;71:103585. https://doi.org/10.1016/j.addma.2023.103585.

[40] Ding D, Pan Z, Cuiuri D, Li H. Wire-feed additive manufacturing of metal components: technologies, developments and future interests. Int J Adv Manuf Technol 2015;81:465–81. https://doi.org/10.1007/s00170-015-7077-3.

[41] Corbin DJ, Nassar AR, Reutzel EW, Beese AM, Kistler NA. Effect of directed energy deposition processing parameters on laser deposited Inconel® 718: External morphology. J Laser Appl 2017;29:22001. https://doi.org/10.2351/1.4977476.

[42] Wang Z, Palmer TA, Beese AM. Effect of processing parameters on microstructure and tensile properties of austenitic stainless steel 304L made by directed energy deposition additive manufacturing. Acta Mater 2016;110:226–35. https://doi.org/10.1016/j.actamat.2016.03.019.

[43] Liu S, Liu W, Kovacevic R. Experimental investigation of laser hot-wire cladding. Proc Inst Mech Eng Part B J Eng Manuf 2015;231:1007–20. https://doi.org/10.1177/0954405415578722.

[44] Cheng B, Lydon J, Cooper K, Cole V, Northrop P, Chou K. Melt pool sensing and size analysis in laser powder-bed metal additive manufacturing. J Manuf Process 2018;32:744–53. https://doi.org/10.1016/j.jmapro.2018.04.002.

[45] Maffia S, Furlan V, Previtali B. Coaxial and synchronous monitoring of molten pool height, area, and temperature in laser metal deposition. Opt Laser Technol 2023;163:109395. https://doi.org/10.1016/j.optlastec.2023.109395.

[46]   Hagan MT, Menhaj MB. Training feedforward networks with the Marquardt algorithm. IEEE Trans Neural Networks 1994;5:989–93. https://doi.org/10.1109/72.329697.

## Vitae

Mostafa Rahmani Dehaghani: Mostafa is a Ph.D. candidate in Mechatronic Systems Engineering at Simon Fraser University, Burnaby, Canada. His areas of expertise are solid mechanics and mechanical design. He is currently working on online modeling and control of metal additive manufacturing processes.

Atieh Sahraeidolatkhaneh: Atieh is a Ph.D. student in Production Systems at University West, Trollhättan, Sweden. Her areas of expertise are robotics and manufacturing engineering. She is currently working on closed-loop control of cooling rate and temperature gradient in metal additive manufacturing processes.

Morgan Nilsen: Morgan is a senior lecturer in Production Systems at University West, Trollhättan, Sweden. He received his Ph.D. in Engineering Science from University West in 2019. His research area is automation with a special focus on sensors, control technology, and process monitoring of welding processes and additive manufacturing.

Fredrik Sikström: Fredrik is an associate professor in Engineering Science at University West, Trollhättan, Sweden. He received his Ph.D. from Chalmers University of Technology in 2019. He is currently focusing on in-process monitoring and control of metallic fusion welding and additive manufacturing using directed energy deposition.

Pouyan Sajadi: Pouyan is a master's student in Mechatronic Systems Engineering at Simon Fraser University, Burnaby, Canada. His areas of expertise are machine learning and regression models. He is currently working on the thermal field prediction of metal additive manufacturing processes using physics-informed neural networks.

Yifan Tang: Yifan is a Ph.D. candidate in Mechatronic Systems Engineering at Simon Fraser University, Burnaby, Canada. He is interested in artificial intelligence-integrated modeling and design optimization methods for expensive engineering problems with data scarcity.

G. Gary Wang: Gary is a professor in Mechatronic Systems Engineering at Simon Fraser University, Burnaby, Canada. He received his Ph.D. from the University of Victoria in Mechanical Engineering in 1999. He has been working on intelligent optimization, design engineering, and product design and development for more than 23 years.